# Stopping powers and energy loss straggling for $(0.9-3.4)$ MeV protons in a kapton polyimide thin film

S. Damache[1], S. Djaroum[2], S. Ouichaoui[3*], L. Amari[3], D. Moussa[3]

[1]*Division de Physique, CRNA, 02 Bd. Frantz Fanon, B.P. 399 Alger-gare, Algiers, Algeria*
[2]*Division de Technologie Nucléaire, CRNB, B.P. 180 Ain-Oussara, Djelfa, Algeria*
[3]*Université des Sciences et Technologie H. Boumediene (USTHB), Faculté de Physique, B.P. 32, 16111 Bab-Ezzouar, Algiers, Algeria*

**Abstract**

The energy loss and energy loss straggling widths have been measured in transmission for $E_p \approx (0.9 - 3.4)$ MeV protons traversing a thin kapton polyimide foil. In a prior step, the thickness and non-uniformity of the target foil were carefully investigated. The overall relative uncertainties in the stopping power and energy loss straggling variance data amount, respectively, to less than 2% and 8%. The $S(E)$ experimental data show to be in excellent agreement with available previous ones and with **those** compiled in the ICRU-49 report. They are fully consistent with the predictions of Sigmund-Schinner's binary collision theory of electronic stopping over the whole proton energy range explored. An average deviation of ~2.5% relative to values calculated by the SRIM-2008 code, likely due to effects of valence electrons involving the $C - H$, $C = C$ and $C = O$ bonds, is however observed at low proton velocities. The measured energy loss straggling data, which are unique to our knowledge, are found to be in good agreement with values derived by the classical Bohr formula for $E_p \gtrsim 1300$ keV but they significantly exceed Bohr's collisional energy loss straggling at lower proton velocities where target electrons can no longer be considered as free. They also show to be consistent with the predictions of the Bethe-Livingston and Sigmund-Schinner theories over the low proton velocity region ($E_p < 1300$ keV). However, they are significantly overestimated by these theories over the intermediate and high proton velocity regions, which may be due to bunching effect by inner shell electrons of the polymer target. Besides, our energy loss straggling data are in better overall consistency with the Yang, O'Connor and Wang empirical formula for $E_p > 1300$ keV, while deviations above the latter amounting up to ~18% are observed at lower proton velocities.

Keywords: Swift protons; Kapton polyimide; Stopping power; Energy loss straggling; Binary collision theory.

*Corresponding author: souichaoui@gmail.com; sdamache@gmail.com

1. **Introduction**

When swift charged particles cross matter, they are slowed down with losing all or part of their energy mainly due to inelastic electronic collisions with target atoms. These collisions are statistical in nature and the resulting energy loss distributions are characterized by the associated mean energy losses and widths (FWHMs related to the energy loss straggling variances). The ion mean energy loss and energy loss straggling variance are statistical quantities of crucial importance regarding both the understanding of the fundamental aspects of ion-matter interactions and ion beam analysis various applications [1-8]. The ion mean energy loss per travelled path length, i.e., the target material stopping power, $S(E)$, has been extensively investigated for various ion-target combinations over a wide range of ion bombarding energies [9-19]. In contrast, the energy loss straggling variance, $\Omega^2(E)$, is much more difficult to measure reliably, and still much more limited corresponding studies are available in the literature [20-25]. Indeed, the energy loss straggling is very sensitive to the target material non-uniformity, roughness and inhomogeneity in chemical composition that may be responsible for undesirable, supplementary broadenings to the effective energy loss straggling variance. These spurious effects of the target material are usually difficult to evaluate experimentally, and special precautions need to be taken for reducing them to a negligible level.

In the present paper, we report on energy loss and energy loss straggling measurements (Section 2) carried out for incident proton energies, $E_P \approx (0.9 - 3.4)$ MeV, crossing a 1062 µg/cm²-thick kapton polyimide foil. The derived stopping power and energy loss straggling data are reported and discussed in Section 3. The $S(E)$ data are compared to previous experimental results available in the literature [26-28] and to values from the ICRU-49 report [29]. They are also discussed in comparison to values calculated by the binary collision approximation scheme (BCAS) of Sigmund and Schinner [30-32] using the last version of the program PASS [33], and to values generated by the SRIM-2008 computer code [34]. The derived energy loss straggling data are compared to the predictions of the Bohr and Bethe-Livingston classical theories [35,36] of collisional energy loss straggling and to values derived by the empirical formula of Yang, O'Connor and Wang [37]. Besides, they are discussed in comparison to values generated by the Sigmund-Schinner [30,38] binary collision theory of electronic stopping using the PASS code [33]. Note that all the calculated $S(E)$ and $\Omega^2(E)$ values have been performed with assuming the validity of the Bragg-Kleeman additivity rule [39] for compound materials. Finally, concluding remarks from the current study are drawn in Section 4.

## 2. Experimental set up and methods

The experimental set up and methods used in this experiment have been described elsewhere [12-17,25]. Then, only an account of the main experimental aspects is given in the following subsection.

### 2.1. Proton beam and detection system

The energy loss and energy loss straggling measurements were carried out at the Algiers 3.75-MV Van de Graaff accelerator whose energy calibration has been checked [16] by exploring narrow resonances in the $^{19}$F(p, αγ)$^{16}$O and $^{27}$Al(p, γ)$^{28}$Si nuclear reactions. A primary proton beam with energy in the range $E_p \approx (0.9 - 3.4)$ MeV, mean current intensity of ~30 nA and circular spot of ~1.5 mm diameter was directed onto a very thin Au-Si scattering target through a series of electrostatic lenses and focusing slits. Then, a secondary proton beam, backscattered off the Au scattering target was used in transmission through the kapton target foil for measuring the proton energy loss distributions. The latter were recorded without and with target foil in place using a system designed [12] to allow one moving the target holder without breaking the high vacuum constantly maintained inside the scattering chamber. The backscattered proton beam was detected at laboratory angle of 165° relative the primary beam direction by means of a 25 mm$^2$ effective area, 500 μm thick ULTRA ion implanted Si detector, collimated by a 3 mm diameter slit and set at ~2 cm from the investigated target foil. A maximum allowed bias voltage of 50 V was applied to the Si detector in order to minimize the associated pulse-high-defect that constitutes the main cause of the detector response non-linearity [40,41]. A data acquisition system made of standard EG & G Ortec electronics was associated to the detector for pulse processing and recording the proton energy loss spectra [14]. The achieved overall energy resolution of the detection system was of ~10.7 keV (FWHM) for 882 keV backscattered protons.

### 2.2. Target foil properties, thickness and non-uniformity investigations

The kapton film was commercially supplied with stated thickness of ~7.5 μm. This polymer compound is characterized by a density of 1.43 g/cm$^3$ and a chemical formula, (C$_{22}$H$_{10}$N$_2$O$_5$)$_n$, involving 31 single bonds (i.e., 13 $C-C$, 6 $C-N$, 2 $C-O$, 10 $C-H$) and 13 double bonds (9 $C=C$, 4 $C=O$) per molecule. Its elemental composition, expressed in weight fraction, is C: 69.12%, H: 2.63%, O: 20.92% and N: 7.33%.

The thickness of the used kapton film was accurately determined using the weight per area method [12]. Circular target samples of ~12 mm in diameter were cut out from the kapton film with a very accurately machined punching tool. They were immersed into an ethanol solution in order to clean them from possible surface dirt. Then, they were dried during ~12 hours in an oven maintained at a temperature of ~50 ° C, and weighted several times each by means of a Sartorius super micro-balance of ± 1µg absolute accuracy. The sample areas were determined by measuring an equivalent one of a ~5 µm-thick nickel foil using a Carl Zeiss Hall 100 optical microscope able to read to ±2 µm. For this purpose, the Ni sample was carefully cut out using the same punching tool as for the kapton samples. A mean areal thickness value, $\bar{x} = (1062 \pm 13)$ µg/cm$^2$, of the target samples was obtained in these measurements, in good agreement with the value given by the supplier.

The macroscopic non-uniformity of the film thickness was estimated via systematic energy loss measurements for 4.824 MeV and 5.485 MeV alpha particles from a very thin $^{241}$Am-$^{239}$Pu-$^{233}$U mixed radioactive source crossing different points over the kapton foil area. This method [12-17, 25] is very powerful for macroscopic non-uniformity scanning [25] since it relates only to $\Delta E_\alpha$ measurements achieved with good accuracy. Multi-Gaussian curves were fitted to the experimental energy loss distributions recorded for alpha particles from the mixed radioactive source crossing the kapton target. This allowed us to determine the alpha particle peak positions with relative uncertainty of less than 0.02%. The foil thickness variations over the scanned points over the target sample area induced fluctuations in the resulting alpha particle $\Delta E_\alpha$ energy loss data lying within one standard deviation of less than 1.5% for both considered alpha particle energies. Finally, a relative uncertainty of ~2% in the determined kapton foil mean thickness was obtained by summing in quadrature the relative uncertainties originating from the target thickness determination (~1.2%) and from the macroscopic non-uniformity scanning (~1.5%).

Furthermore, at a microscopic scale the non-uniformity of the kapton foil was investigated similarly as in [25,42] via energy loss and energy loss straggling measurements for a 2.4 MeV alpha particle beam from the V.d.G. accelerator crossing the kapton foil. The energy loss distributions recorded with and without exposing the kapton foil to the alpha particle beam were adjusted by Gaussian fits allowing a precise determination of the corresponding mean peak positions and FWHM widths. Then, the energy loss straggling variance data, $\Omega^2 = \langle(\Delta E - \langle \Delta E \rangle)^2\rangle$, can be elementarily deduced from the determined widths (FWHM) of these Gaussian shaped energy loss distributions. Finally, an upper bound value of the kapton foil roughness, $\rho^2 < \frac{\Omega^2}{\langle \Delta E \rangle^2} = 3.17 \times 10^{-4}$, was deduced from the derived results of the squared mean energy loss, $\langle \Delta E \rangle^2$, and straggling variance, $\Omega^2$.

### 2.3. Energy loss and energy loss straggling measurements

The experimental data of energy loss and energy loss straggling widths, $\Delta E$ and $\delta E_{exp}$, were respectively deduced from the peak positions and widths (FWHM) of the measured proton energy loss distributions. Similarly as in [12,16] the latter have been carefully inspected for their Gaussian shapes by evaluating the overall dimensionless path-length [43,44] and distribution $\gamma_3^2$-skewness [12,45,46] parameters both for small (3% $\lesssim \frac{\Delta E}{E} \lesssim$ 5%,) and large (20% < $\frac{\Delta E}{E}$ < 36%) energy loss fractions, respectively. The energy loss distributions recorded without and with the kapton foil exposed to a beam of 900 keV protons are reported in Fig. 1. These energy spectra correspond to the extreme case of maximum energy loss fraction, $\frac{\Delta E}{E} = 35.5\%$, considered in this work. Generated Gaussian fits to these experimental data are also plotted in Fig. 1 (solid curves), allowing an accurate determination of the distribution peak positions and widths. As can be seen, the measured spectra are well reproduced by the Gaussian fits over the considered channel intervals defined by ~3 up to 4 times one standard deviation around the peak positions [12]. The difference in mean peak positions in the recorded proton energy loss spectra without and with the kapton target in place (see Fig.1) yields the mean energy loss, $\Delta E$, while the energy loss straggling width, $\delta E_{exp}$, was deduced from the measured widths (FWHM), $\delta E_0$ and $\delta E_{st}$, of the unstraggled and straggled beam profiles using the relation:

$$\delta E_{exp}^2 = \delta E_{st}^2 - \delta E_0^2. \qquad (1)$$

### 2.4. Stopping power and energy loss straggling variance determinations

The experimental stopping power values were deduced from the mean energy loss data as in [12, 14]. Briefly, for energy loss fractions, $\frac{\Delta E}{E} < 20\%$, the average stopping power of the target sample, $S(\bar{E})$, at the proton mean energy, $\bar{E} = E - \frac{\Delta E}{E}$, was derived as the ratio, $\frac{\Delta E}{\bar{x}}$, with 0.05% accuracy [47]. Note that here $E$ is the energy of backscattered protons off the Au-Si scattering target. In some cases where large energy losses were recorded (Table 1), two corrections have been considered: (i) a quadratic correction term derived from an expansion of the $S(\bar{E})$ function as in reference [47] and (ii) the increase in the projectile path length within the kapton target foil due to multiple scattering [12,48,49]. However, these two corrections amounted to at most ~0.17% and ~0.06%, respectively. For energy loss distributions of Gaussian shapes, the measured straggling width, $\delta E_{exp}$, is elementarily related to the experimental energy straggling variance, $\Omega_{exp}^2$, by the expression

$$\delta E_{exp}^2 = 8ln(2)\Omega_{exp}^2. \qquad (2)$$

The present energy loss straggling experimental data were corrected for target foil roughness and inhomogeneity effects using the method due to Besenbacher et al. [50]. Then, the corrected straggling variance is given by

$$\Omega^2(\bar{E}) = \Omega_{exp}^2(\bar{E}) - \rho^2 \langle \Delta E \rangle^2, \qquad (3)$$

where $\langle \Delta E \rangle$ is the ion experimental mean energy loss in the target.

## 3. Results and discussions

The obtained results of the stopping power, $S(\bar{E})$, and energy loss straggling variance, $\Omega^2(\bar{E})$, for protons in kapton polyimide are reported in Table 1 together with the corresponding proton energies and energy loss fractions. The overall relative uncertainties in these $S(\bar{E})$ and $\Omega^2(\bar{E})$ experimental data, evaluated as in [12,25], amount to at most 2% and 8%, respectively. The $S(\bar{E})$ experimental data are compared in Figs. 2 to other data sets from previous experiments [26-28] and from the ICRU-49 report [29]. In Fig. 3, our data are compared to $S(E)$ values calculated by the binary collision approximation scheme (see [30-32] and references therein) using the last version of the PASS program [33] or generated by the SRIM-2008 code [34].

Concerning the energy loss straggling, our experimental data are plotted in Fig. 4 in form of the variance ratio, $\frac{\Omega^2}{\Omega_B^2}$, versus the proton mean energy with $\Omega_B^2 = 85.47$ keV$^2$ being the collisional energy loss straggling variance calculated by Bohr's theory [35] for a 1062 µg/cm²-thick kapton foil assuming the validity of the Bragg-Kleeman additivity rule. Observing the trend of the energy loss straggling data over the low proton velocity region, $E_p$ < 1300 keV, we have noted an apparently large departure between the two data points at 950 keV and 1050 keV, lying well outside the initially calculated error bars. This has led us to carefully reexamine the corresponding energy spectra with and without the kapton target being exposed to the proton beam. Then, plotting the widths (FWHM) of the measured energy loss distribution versus the mean proton energy for these two data points and their immediate neighbors, we have pointed out a presumably pronounced fluctuation of the proton beam spot width with the kapton target being placed (for the point at ~950 keV) and without target in place (for the point at 1050 keV). Consequently, we have reevaluated the relative uncertainties in the latter two data points that have been increased, respectively, from 2.98% up to 7.86% and from 2.59% up to 5.01%, as can be seen in Fig. 4. In the latter representation, the measured energy loss straggling data are also conveniently compared to the predictions of: (i) the Bethe-Livingston theory [36] for the collisional energy loss straggling assuming the target electrons to be bound within the atomic shells, (ii) Sigmund-Schinner's binary collision theory [30,38] using the PASS computer code [33], and (iii) Yang et al. [37] empirical formula. In the Bethe-Livingston theory, the effective target atomic number expresses as, $Z'_2 = Z_2 - \sum_i Z_i$, where the sum extends over the electrons excluded by the condition, $I_i > 2mv^2$, with $m$ and $v$ denoting the electron mass and projectile velocity, respectively. Note that in all the energy loss straggling calculations the input data used for the H, C, N and O constituent elements of the kapton target, reported in Tab.2, were taken from the ICRU-73 report [51].

**3.1. Stopping power data**

As can be seen in Fig. 2, our $S(\bar{E})$ experimental results are in fairly good agreement with both previous data from reference [26] and the ICRU-49 report [29] over the common explored proton energy ranges. They are also in excellent agreement with experimental data from references [27,28] for $E_p < 1100$ keV. However, they lie above the data from reference [28] by ~3 up to 6% and slightly below the data from reference [27] over the common explored higher energy interval. Relative to the $S(E)$ values derived by the SRIM-2008 [34] and the PASS [33] computer codes based on the binary collision approximation (see Fig.3), our data show to be fully consistent with the BCAS predicted values [30,31,32] over the entire proton energy range explored, and in good agreement (within the experimental uncertainties) with values derived by the SRIM-2008 code [34] over the proton energy region, $E_p \sim (785 - 1300)$ keV. In the remaining proton energy regions explored, our $S(\bar{E})$ data appear to be slightly overestimated by the SRIM-2008 code at higher energies ($E_p > 1300$ keV) while they are underestimated by the latter computer code at lower proton energies ($E_p < 785$ keV). An average deviation of ~2.5% is clearly observed between our data and SRIM-2008 calculated values in both precedent energy ranges. Notice that these trends of the experimental stopping powers of compounds relative to the values calculated by the SRIM code - indicating that more pronounced deviations are observed for oxides, nitrides and hydrocarbons - have also been pointed out previously by several groups of authors [12,14,52-56].

## 3.2. Energy loss straggling data

As can be seen in Fig. 4, the energy loss straggling ratio, $\frac{\Omega^2}{\Omega_B^2}$, remains essentially constant in overall scale and close to unity for proton energies, $E_p > 1300$ keV, while in the remaining lower energy region explored this ratio increases with decreasing the proton energy up to a maximum value of ~1.33 around $E_p = 1000$ keV. The present energy loss straggling data are consistently reproduced, in average, by the classical Bohr theory [35] of collisional energy straggling over the high proton energy region explored, i.e., for $E_p \gtrsim 1300$ keV, where the bindings of target electrons within atomic shells are neglected. In contrast, our data are substantially underestimated by Bohr's theory over the lower proton energy region where the observed deviation amounts up to ~22%, in average. Thus, a general trend of the energy loss straggling experimental data is that they exhibit significant energy dependence over the low proton energy region below $E_p \approx 1300$ keV where target electrons are increasingly lesser free and subject to binding effects with decreasing the proton energy. Similar such trends of the energy loss straggling for $(0.5 < E_p < 2)$ MeV protons transmitted through compound targets have been observed previously for organic target materials of various thicknesses [25,52,57]. Besides, one can observe in Fig. 4 that our energy loss straggling data tend to the values predicted by the Bethe-Livingston [36] and Sigmund-Schinner [30,38] theories over the low proton energy range explored, $E_p < 1300$ keV, while they are significantly overestimated by the latter two theories for higher proton energies. Within the intermediate proton energy region, $1300 < E_p \lesssim 2500$ keV, where the measured energy loss distributions are fully Gaussian-shaped, the observed discrepancies with the predictions of these theories vary between 11% and 17%: they are likely due to phase and spatial correlation effects of inner shell atomic electrons of the studied kapton target. Indeed, the input data for the H, O and N constituents of the latter used in our energy loss straggling calculation by the Bethe-Livingston theory [36] are those for molecular gases. Moreover, the bunching of electrons in their atomic orbits was ignored in this calculation, which may lead to a decrease of the energy loss straggling as predicted by this theory [36] for solid targets.

In our calculation of the energy loss straggling with the Yang et al. [37] empirical formula, we have also assumed the (H, N, O) elemental constituents of the kapton polyimide target to be solids. However, phase state effects, i.e., the solid-gas difference, has practically no incidence on the Yang et al. calculated values over the whole considered proton energy range. The induced difference in the energy loss straggling variances amounts only to less than 2%. Thus, our energy loss straggling experimental data can be considered in overall agreement with the values derived by the Yang et al. empirical formula [37] for $E_p > 1300$ keV, while for lower proton energies they lie significantly higher than estimated with this formula. The observed deviation, that increases as the proton energy decreases towards the stopping power function maximum, amounts up to ~18%. Similar large deviations between experimental data and the energy loss straggling derived by the Yang et al. empirical formula have also been recently reported [57] for C, O, Al and Kr ions crossing $ZrO_2$ targets. Notice, however, that in addition to assuming an ionic charge of unity for the incident hydrogen ions, Yang et al. [37] have fitted a deformed resonant function to experimental data affected by discrepancies of up to 60%.

## 4. Summary and conclusion

The observed general consistencies between theoretical predictions and our stopping power and energy loss straggling experimental data over the explored proton velocity regions clearly attest for the reliability of our measurements.

The measured $S(\bar{E})$ stopping power data for ~(0.9 – 3.4) MeV protons crossing a thin kapton polyimide foil are found in overall agreement with previously measured data [26-28] and with the ICRU-49 compilation. As discussed in the preceding section, they exhibit similar behaviors versus the proton energy relative to $S(E)$ values calculated by theory. Notably, they are fully consistent with the predictions of Sigmund-Schinner's binary collision theory of electronic stopping over the whole proton energy range explored. Relative to values derived by the SRIM-2008 code based on the binary collision approximation, a good agreement is observed over the proton energy region, $E_p \sim (785 - 1300)$ keV, while expected, small, deviations are pointed out, as has been observed in several previous studies.

Concerning the energy loss straggling experimental data for the same projectile-target system, one notes that they exhibit marked energy dependencies with two main distinct trends (see Fig. 4). (i) Over the high proton velocity region explored, i.e., for $E_p > 1300$ keV, they are in overall good agreement with Bohr's classical theory predicting energy-independent collisional energy loss straggling with neglecting the bindings of the target atomic electrons. In this region, however, our data are slightly overestimated by the Bethe-Livingston theory assuming the target electrons to be bound within their atomic shells and by Sigmund-Schinner's binary collision theory involving both the shell and Barkas-Andersen corrections. (ii) Over the lower proton velocity region explored here, i.e., for $E_p < 1300$ keV, our data significantly increase as the proton energy decreases towards the stopping power function maximum where complex interaction effects, including charge exchange and target-projectile correlation effects, are expected to take place. Although the experimental data exhibit a more abrupt enhancement towards low proton velocities, this trend is qualitatively consistent with the energy loss straggling variations predicted both by the Bethe-Livingston and Sigmund-Schinner theories. However, none of the performed theoretical calculations does completely account for the observed trends of the energy loss straggling experimental data throughout the pointed out proton velocity regimes, especially the pronounced enhancement of the data at low projectile velocities. It turns out, consequently, that together with additional accurate experimental data, theoretical refinements, such as the inclusion of projectile-target correlation effects like the electron bunching and atomic packing, are still needed in order to better clarify the situation within the low projectile velocity region. Besides, even the Yang, O'Connor and Wang empirical formula that, apart from Bohr's theory, yields the closest energy loss straggling values to experimental data over the high proton velocity regime, appears to be obviously exceeded over the low projectile velocity regime. It therefore needs to be further improved with considering up-to-date and accurate experimental data.

**Tables:**

| $\bar{E}$ (keV) | $\frac{\Delta E}{E}$ (%) | $S(\bar{E})$ (keV) | $\Omega^2$ (keV$^2$) |
|---|---|---|---|
| 725.4  | 35.52 | 294.99 ± 5.91 | 104.60 ± 2.65 |
| 785.6  | 31.25 | 273.96 ± 5.49 | 104.11 ± 2.53 |
| 840.8  | 28.43 | 262.35 ± 5.25 | 109.69 ± 3.03 |
| 948.9  | 23.96 | 243.19 ± 4.87 | 099.46 ± 2.96 |
| 1055.9 | 20.45 | 226.43 ± 4.54 | 113.91 ± 2.95 |
| 1161.4 | 17.70 | 212.32 ± 4.25 | 104.30 ± 2.24 |
| 1265.4 | 15.55 | 200.91 ± 4.02 | 094.88 ± 3.10 |
| 1370.3 | 13.58 | 187.96 ± 3.77 | 084.28 ± 3.61 |
| 1472.9 | 12.15 | 179.42 ± 3.59 | 090.13 ± 3.20 |
| 1575.6 | 10.86 | 170.44 ± 3.41 | 088.22 ± 2.73 |
| 1677.8 | 09.79 | 162.63 ± 3.26 | 090.36 ± 3.00 |
| 1779.4 | 08.89 | 155.90 ± 3.12 | 086.73 ± 3.29 |
| 1880.7 | 08.11 | 149.63 ± 3.00 | 080.05 ± 3.42 |
| 2082.5 | 6.83  | 138.71 ± 2.78 | 082.98 ± 4.22 |
| 2282.8 | 5.90  | 130.63 ± 2.62 | 082.76 ± 3.42 |
| 2483.6 | 5.07  | 121.60 ± 2.44 | 082.27 ± 4.20 |
| 2683.8 | 4.40  | 113.82 ± 2.28 | 091.11 ± 4.06 |
| 2881.9 | 3.97  | 109.80 ± 2.20 | 089.02 ± 4.47 |
| 3081.4 | 3.50  | 103.22 ± 2.07 | 088.78 ± 4.30 |
| 3279.6 | 3.16  | 099.11 ± 1.99 | 082.93 ± 3.94 |

**Table 1:** Measured stopping power data (column 3) and energy loss straggling data (column 4) versus the proton mean energy (column 1) for the used kapton polyimide target. The corresponding energy loss fractions are also given in column 2.

| Element | $Z_2$ | Subshell $i$ | Subshell occupation $Z_2 f_i$ | Subshell $I$-value $\hbar\omega_j$ (eV) | Subshell binding energy $U_i$ (eV) |
|---|---|---|---|---|---|
| H | 1 | K: $n=1; l=0$ | 1 | 19.2 | 15.42 |
| C | 6 | K: $n=1; l=0$ | 1.992 | 486.2 | 288.2 |
|   |   | L$_I$: $n=2; l=0$ | 1.841 | 60.95 | 16.59 |
|   |   | L$_{II}$: $n=2; l=1$ | 2.167 | 23.43 | 11.26 |
| N | 7 | K: $n=1; l=0$ | 1.741 | 732.61 | 403.8 |
|   |   | L$_I$: $n=2; l=0$ | 1.680 | 100.646 | 20.33 |
|   |   | L$_{II}$: $n=2; l=1$ | 3.579 | 23.55 | 14.534 |
| O | 8 | K: $n=1; l=0$ | 1.802 | 965.1 | 538.2 |
|   |   | L$_I$: $n=2; l=0$ | 1.849 | 129.85 | 28.7 |
|   |   | L$_{II}$: $n=2; l=1$ | 4.349 | 31.60 | 13.618 |

$n$ and $l$ are respectively the principal and azimuthal quantum numbers.

**Table 2:** Average excitation energy, $I_i$, electron number, $Z_i = f_i Z_2$ (where $f_i$ is the oscillator-strength of the ith shell/subshell) and binding energy, $U_j$, for shells or subshells of the (H, C,

N et O) elementary constituents of the studied kapton polyimide target. These input data used in our calculations were extracted from the ICRU-73 report [51].

**Figures:**

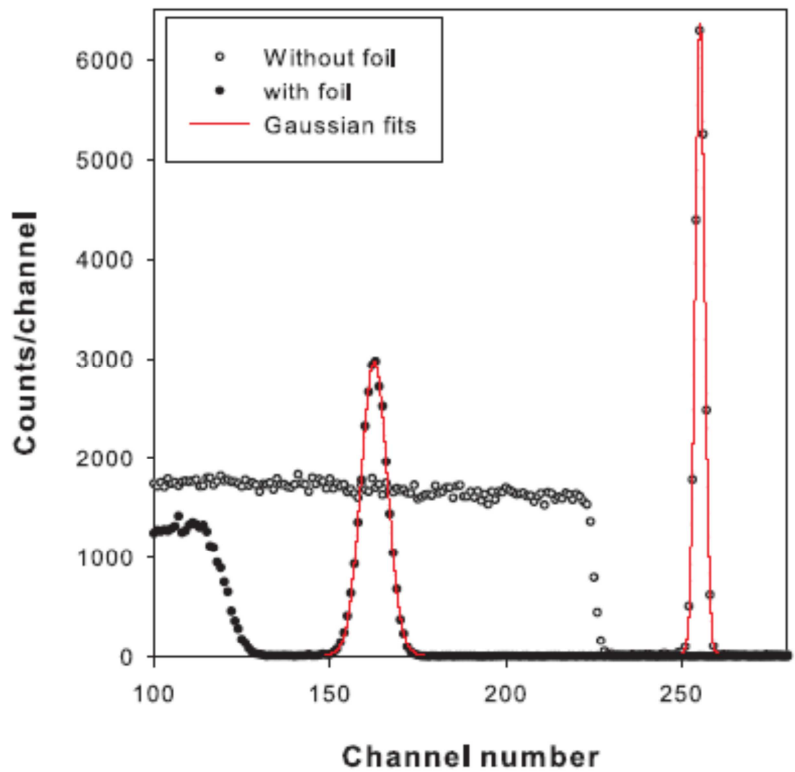

**Fig. 1:** Typical energy loss spectra recorded with and without the kapton foil in place for 900 keV incident protons using a 2048-channel portion of an Ortec MCB Card.

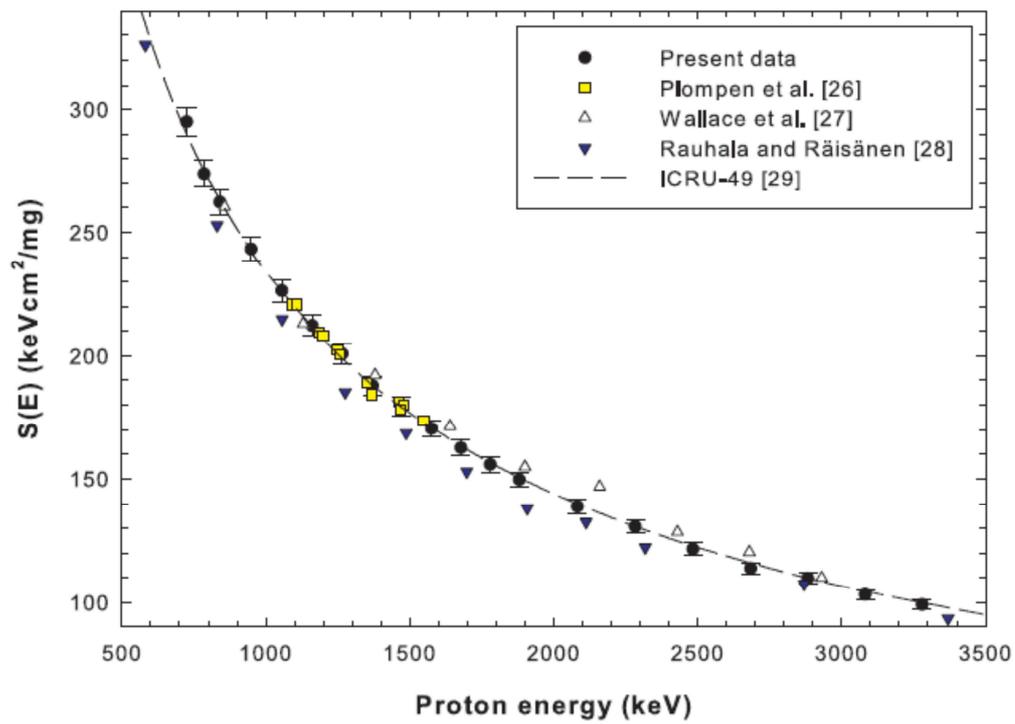

**Fig. 2:** Measured stopping powers of kapton polyimide versus the proton mean energy compared to previous $S(E)$ experimental data [26,27,28] and to values compiled in the ICRU-49 report [29].

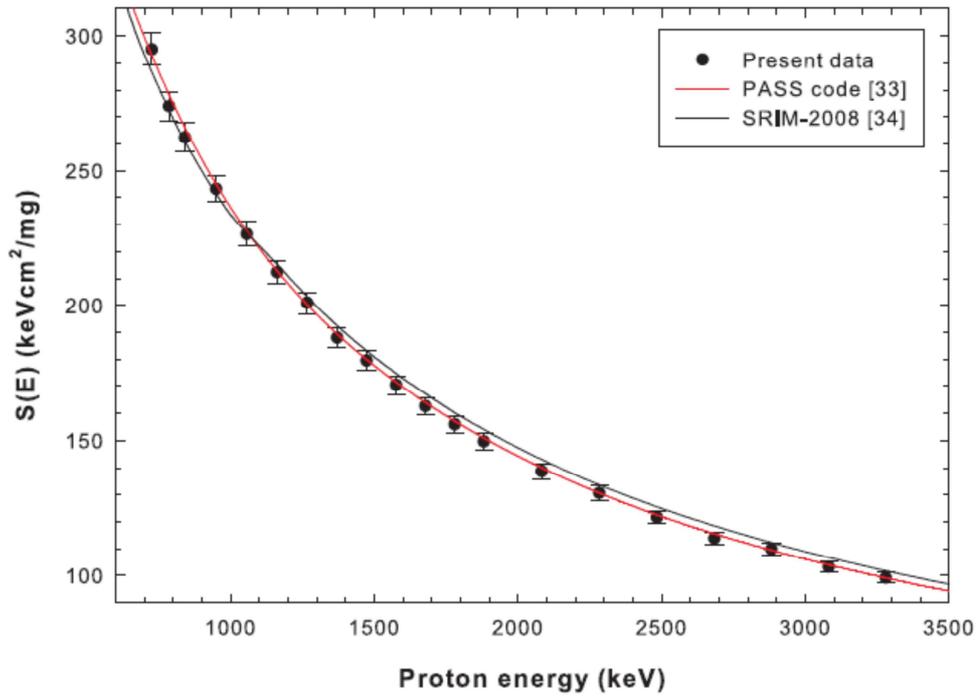

**Fig. 3:** Measured stopping powers of kapton polyimide versus the proton mean energy compared to $S(E)$ values calculated by the Sigmund-Schinner binary collision approximation scheme (BCAS [30, 31,32]) using the last version of program PASS [33] and to values generated by the SRIM-2008 computer code [34].

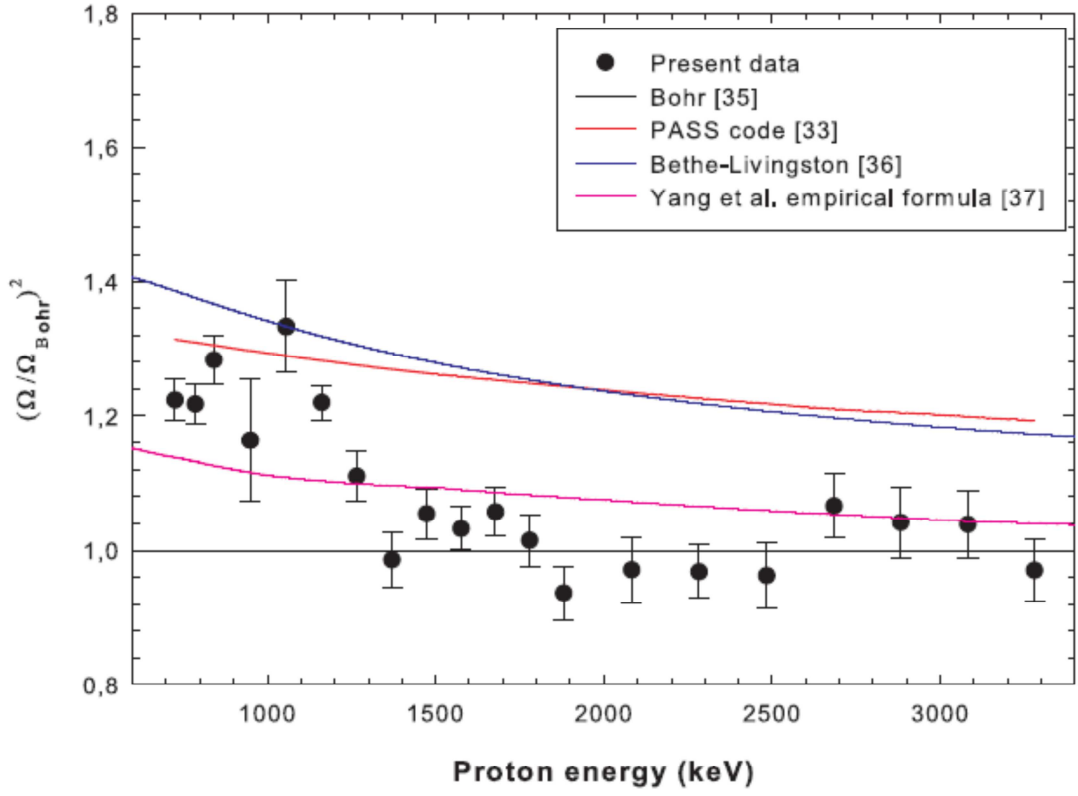

**Fig. 4:** Measured energy loss straggling data relative to Bohr's classical theory [35] of collisional energy loss straggling (variance ratio, $\frac{\Omega^2}{\Omega_B^2}$) for incident protons in kapton polyimide versus the mean proton energy compared to values predicted by the Bethe-Livingston [36] and Sigmund-Schinner [30,31,32] (PASS calculation [33]) theories, and by Yang et al. empirical formulae [37] assuming Bragg-Kleemann's additivity rule for compound materials.